\documentclass[conference,anonymous]{IEEEtran}
\IEEEoverridecommandlockouts
\usepackage{cite}
\usepackage{verbatim}
\usepackage{amsmath,amssymb,amsfonts}
\usepackage{algorithmic}
\usepackage{graphicx}
\usepackage{textcomp}
\usepackage{xcolor,colortbl}
\def\BibTeX{{\rm B\kern-.05em{\sc i\kern-.025em b}\kern-.08em
    T\kern-.1667em\lower.7ex\hbox{E}\kern-.125emX}}
    
\makeatletter
\newcommand{\linebreakand}{%
  \end{@IEEEauthorhalign}
  \hfill\mbox{}\par
  \mbox{}\hfill\begin{@IEEEauthorhalign}
}
\makeatother

\begin{document}

\title{Supporting AI Engineering on the IoT Edge through Model-Driven TinyML
}

\author{\IEEEauthorblockN{Armin Moin}
\IEEEauthorblockA{\textit{Department of Informatics} \\
\textit{Technical University of Munich, Germany} \\
moin@in.tum.de}
\and
\IEEEauthorblockN{Moharram Challenger}
\IEEEauthorblockA{\textit{Department of Computer Science} \\
\textit{Univ. of Antwerp \& Flanders Make, Belgium} \\
moharram.challenger@uantwerpen.be}
\linebreakand
\IEEEauthorblockN{Atta Badii}
\IEEEauthorblockA{\textit{Department of Computer Science} \\
\textit{Univ. of Reading, United Kingdom} \\
atta.badii@reading.ac.uk}
\and
\IEEEauthorblockN{Stephan G{\"u}nnemann}
\IEEEauthorblockA{\textit{Dep. of Informatics \& Munich Data Science Institute} \\
\textit{Technical University of Munich, Germany} \\
guennemann@in.tum.de}
}
\maketitle

\begin{abstract}
Software engineering of network-centric Artificial Intelligence (AI) and Internet of Things (IoT) enabled Cyber-Physical Systems (CPS) and services, involves complex design and validation challenges. In this paper, we propose a novel approach, based on the model-driven software engineering paradigm, in particular the domain-specific modeling methodology. We focus on a sub-discipline of AI, namely Machine Learning (ML) and propose the delegation of data analytics and ML to the IoT edge. This way, we may increase the service quality of ML, for example, its availability and performance, regardless of the network conditions, as well as maintaining the privacy, security and sustainability. We let practitioners assign ML tasks to heterogeneous edge devices, including highly resource-constrained embedded microcontrollers with main memories in the order of Kilobytes, and energy consumption in the order of milliwatts. This is known as TinyML. Furthermore, we show how software models with different levels of abstraction, namely platform-independent and platform-specific models can be used in the software development process. Finally, we validate the proposed approach using a case study addressing the predictive maintenance of a hydraulics system with various networked sensors and actuators.
\end{abstract}


\section{Introduction}\label{introduction}
Finding IT professionals, e.g., software developers and system engineers who are familiar with the diverse hardware and software platforms, programming languages, communication protocols and APIs that are involved in the Internet of Things (IoT) applications is very difficult. The technologies are diverse and the platforms have a broad spectrum, ranging from tiny sensors and microcontrollers with a few Kilobytes (KB) of main memory and very constrained power and processing resources to capable cloud servers with multiple GPUs and large in-memory databases. Consequently, their operating systems (if any), programming languages and communication protocols are also very different. For instance, a single IoT project might require familiarity with various operating systems, such as Linux, TinyOS, ContikiOS, ROS, Android, iOS as well as machine codes (machine languages) or assembly languages of multiple microcontrollers and computers with various architectures. Moreover, different programming languages will be used for different parts of the software systems that need to be deployed on distributed platforms. C, Java (J2SE and J2EE), Python, PHP, Go and Javascript are only a few examples of the common choices. In addition, communication protocols are diverse on different layers of the network stack. For example, on the application layer, CoAP (Constrained Application Protocol) and MQTT (Message Queuing Telemetry Transport) are more suitable protocols than HTTP (Hypertext Transfer Protocol) for resource-constrained devices. The former (CoAP) is suitable for one-to-one communications, whereas the latter (MQTT) is designed for many-to-many communications following the publish-subscribe pattern. Therefore, no matter how professional and skilled a software developer or system engineer is, they can neither master the entire technology spectrum and cross-domain functions of a complex IoT system nor can they efficiently communicate and collaborate with other experts working on the same project.

The Model-Driven Software Engineering (MDSE) paradigm, also known as Model-Based Software Engineering (MBSE), specifically the Domain-Specific Modeling (DSM) methodology with full code generation \cite{KellyTolvanen2008} offers abstraction and automation to deal with the above-mentioned complexity. Over the past decade, its applications have been expanded from the niche domains, such as embedded systems for safety critical applications, e.g., in the automotive industry, to the more complex domains, such as the Internet of Things (IoT) \cite{Harrand+2016} with highly heterogeneous, distributed systems of systems, called Cyber-Physical Systems (CPS) \cite{GeisbergerBroy2014, Schaetz2014}. Examples of tools offering solutions for domain-specific MDSE of embedded systems comprise MATLAB Simulink \cite{Simulink} and AutoFOCUS \cite{Aravantinos+2015}. Moreover, ThingML \cite{Morin+2017, Harrand+2016, Fleurey+2011, ThingML}, HEADS \cite{Morin+2016, HEADS} and $\mu$-Kevoree \cite{Fouquet+2012} supported domain-specific MDSE for the IoT. These solutions enabled full source code generation in an automated manner. 

However, today's software systems that support IoT services require more smart capabilities, as well as more dynamicity at runtime. There is a trend towards data-driven software and systems design and modeling. This handles uncertainty at the software design-time. In other words, the modeler or designer of an IoT service that requires AI, or more specifically ML, may postpone certain design-time decisions and leave them to the predictions, recommendations or outputs of trained ML models at the runtime. Previous work exists in the literature, e.g., ML-Quadrat \cite{Moin+2022-SoSyM, ML-Quadrat} that introduced this. Nevertheless, their major drawback was that they did not support deploying ML models on the IoT edge devices. In contrast, for many IoT use cases today, the deployment of compact ML models that can function self-sufficiently on resource-constrained microcontrollers is necessary.

For instance, a modern smartphone that runs an iOS or Android operating system has a Digital Signal Processor (DSP) chip with power consumption in the order of only a few milliwatts (mW). This microprocessor chip has the task of continuously listening to the surrounding environment for a predefined statement in the form of a speech command, i.e., a so-called \textit{wake word}, such as \textit{\lq{}Hey Siri\rq{}} or \textit{\lq{}OK Google\rq{}}. Note that this always-on DSP must be separated and independent of the main CPU so that the smartphone can save the battery by letting its main CPU that consumes considerably more energy stand by in an inactive mode as long as possible. The trained Artificial Neural Network (ANN) ML model that can efficiently perform the mentioned task on such a DSP has a size of about 14 KB. 
There are many further scenarios beyond the above-mentioned example. For instance, Park et al. \cite{Park+2018} proposed an ANN model with the size of around 15 MB for enhanced, real-time, automatic speech recognition on smartphones and embedded devices. In contrast to the previous case, where speech recognition was only needed for a simple wake word, this model is much more capable in terms of speech recognition. However, the advanced capability of this model comes at a cost: Its size is more than one thousand times larger than the aforementioned wake word recognizer ANN model. Thus, it cannot fit into the memory of typical TinyML devices. 

Furthermore, in other IoT use cases, such as predictive maintenance, the so-called \textit{peel-and-stick sensors}, which require no battery change over their lifetime, or the \textit{Everactive wireless sensors}, which are exclusively powered by low levels of harvested energy from the surrounding environment are increasingly becoming prevalent. These tiny IoT sensors are either already AI-enabled or are expected to become so in the future \cite{WardenSitunayake2019}.

In this paper, we propose a novel approach to domain-specific MDSE of smart services that will run on heterogeneous and distributed IoT devices. In particular, we advocate edge computing (fog computing), specifically edge analytics and TinyML. As mentioned, the latter involves delegating ML tasks to highly resource-constrained microcontrollers with ultra-low energy consumption levels. This lets the data remain at the edge of the network (i.e., on the users' side) and be processed there instead of being transferred to the other nodes of the distributed system, e.g., to the cloud. One of the main drivers for this paradigm shift at the present time is of privacy concerns and the need to ensure legal compliance assurance by design, e.g., concerning the General Data Protection Regulation (GDPR) in the European Union (EU) and the California Consumer Privacy Act (CCPA) in the United States (US) \cite{Li+2021}. In addition, the device energy efficiency, the network throughput optimization and the availability of the service or certain functionalities irrespective of the network conditions will be other benefits of the said transition regarding the execution of data analytics and ML on the edge devices.

Our main Research Questions (RQ) are the following: \textbf{RQ1.} Can we enable automated full code generation out of the software models of smart IoT services that will deploy trained ML models on highly resource-constrained IoT edge platforms (i.e., realizing TinyML)? \textbf{RQ2.} Can we have a higher level of abstraction for the Platform-Independent Models (PIM) that will abstract from the details and constraints of the underlying IoT platforms, and simultaneously a lower level of abstraction for the Platform-Specific Models (PSM) out of which the full implementation must be generated? Note that PIM and PSM here refer to the software and system models, not the data analytics and ML models. Hence, the contribution of this paper is two-fold: (i) It assesses RQ1 and enables TinyML in the domain-specific MDSE of smart IoT services. (ii) It assesses RQ2 and provides the PIM and the PSM layers for the software models of smart IoT services.

The rest of this paper is structured as follows: Section \ref{background} offers some background information. Then, we review the state of the art briefly in Section \ref{sota}. We propose our novel approach in Section \ref{proposed-approach}. Further, Section \ref{validation} implements and validates the proposed approach. Finally, we conclude and suggest future work in Section~\ref{conclusion-future-work}.

\section{Background}\label{background}
In this section, we provide some required background information on MDSE and TinyML.

\subsection{MDSE}\label{background-mdse}
There exist different approaches to the MDSE paradigm. The Model-Driven Architecture (MDA) standard of the Object Management Group (OMG) \cite{OMGMDA2014}, which was initially issued in 2001 and then updated in 2014, serves as a key reference for MDSE. According to MDA, three default architecture viewpoints are defined for every system: computation-independent, platform-independent and platform-specific. Computation-Independent Models (CIM) are business or problem-domain models. They use the vocabulary that is familiar to the subject matter experts in the respective domains. However, Platform-Independent Models (PIM) are solution-domain models, namely models that are related to the computational concepts. Nevertheless, they abstract from the details of any specific platform. Further, Platform-Specific Models (PSM) augment PIMs with the details that are specific to particular platforms. In MDA, the requirements specified in a CIM must be traceable to the constructs in the PIM and the PSMs that implement them (and vice-versa) \cite{OMGCephas2006}.

However, efficient full code generation that does not require any further manual development is often not feasible with MDA. MDA is rather generic and broad. Moreover, the round-tripping processes (i.e., the model transformations from the CIM to the PIM and then the PSMs and vice-versa) result in many model artifacts that need to be managed and might not be consistent over the time. In contrast, the Domain-Specific Modeling (DSM) methodology \cite{KellyTolvanen2008} that is adopted and adapted in this work promotes narrowing the domain of interest down and also avoiding round-tripping. Models are very specific to a particular use case and the full implementation of the solution is generated out of the models, namely PSMs in the MDA terminology. In this paper, we use both PIM and PSM, but avoid round-tripping. A PSM in our software development methodology is simply an extension of a PIM with the platform-specific \textit{annotations} and \textit{configurations} that are necessary for the code generation for a certain target IoT platform out of the PSM. By \textit{a particular platform}, we mean the combination of the hardware architecture, the operating system if applicable, the programming language, the libraries and APIs, as well as the communication protocols.

\subsection{TinyML}\label{background-tinyml}
Running Machine Learning (ML) tasks, such as making predictions using ML models on embedded devices is a young field. In the context of the IoT, i.e., for the networked embedded devices, this is inline with the trend towards assigning more computational tasks to the edge of the network as opposed to the cloud. The trend is known as edge computing or fog computing. The IoT edge devices reside on the users' side and can range from desktop PCs, laptops, tablets and smartphones to embedded single-board computers, such as Raspberry Pi and embedded microcontrollers. In the case of deploying pre-trained ML models on the resource-constrained microcontrollers with ultra-low power consumption in the range of 1 mW, we are dealing with TinyML. These microcontrollers possess main memories (RAM) in the order of tens to hundreds of kilobytes. In addition, their persistent flash memories can be in the order of kilobytes to megabytes. Moreover, their CPU clock speeds might be as low as just tens of MHz. Hence, they are small and highly energy efficient. Last but not least, they are relatively inexpensive and can be ordered in large quantities. 

In this work, we use an \textit{Arduino Nano 33 BLE Sense} microcontroller board \cite{ArduinoNano33BleSense} with an ARM\textsuperscript{\textregistered} Cortex\textsuperscript{\textregistered}-M4 32-bit processor with a clock speed of 64 MHz, 256 KB RAM, 1 MB flash memory and various on-board sensors, e.g., for the temperature, humidity, pressure, brightness, vibration, etc. for the TinyML platform. In contrast, another target platform for code generation will be an embedded single-board computer, namely a \textit{Raspberry Pi 3 B+} board \cite{RaspberryPi3B+} with an 
ARM\textsuperscript{\textregistered} Cortex-A53 (ARMv8) 64-bit SoC (System on Chip) that has a clock speed of 1.4 GHZ, as well as 1 GB of main memory. Although this does not fall under the category of the TinyML platforms, it will be used to demonstrate the heterogeneity of the target platforms for the fully-automated code generation out of the software models.

\section{State of the Art}\label{sota}
As set out in Section \ref{introduction}, ThingML \cite{Morin+2017, Harrand+2016, Fleurey+2011, ThingML} and HEADS \cite{Morin+2016, HEADS} supported the MDSE paradigm, specifically the DSM methodology \cite{KellyTolvanen2008} for full code generation in the IoT/CPS domain. They were based on the Eclipse Modeling Framework (EMF) \cite{emf} and the Xtext framework \cite{xtext}. While they mainly focused on the design-time of software systems, other approaches, such as $\mu$-Kevoree \cite{Fouquet+2012} concentrated on \textit{Models@Runtime}, thus conflating the two distinct phases of design (modeling) and execution of IoT services. The major shortcoming of all the said approaches was the lack of DAML support at the modeling level. In other words, the users of those DSMLs might not deploy the APIs of DAML libraries and frameworks in their software models. Hence, there was no seamless integration between the software models and the DAML models. 

GreyCat \cite{GreyCat} by Hartmann et al. \cite{Hartmann+2017, Hartmann+2018, Hartmann+2019} and ML-Quadrat \cite{ML-Quadrat} by Moin et. al \cite{Moin+2022-SoSyM, Moin+2020, Moin+2018} filled this gap in. However, they fell short of supporting edge analytics and TinyML on resource-constrained IoT devices. The former, which was based on the Kevoree Modeling Framework (KMF) \cite{Francois+2014, kmf} and $\mu$-Kevoree \cite{Fouquet+2012} could generate Java and Javascript/Typescript code. This was not sufficient for many IoT use case scenarios that involved resource-constrained devices that were incapable of executing the Java Virtual Machine (JVM) for the backend. Finally, the latter offered code generation for a range of platforms, including C code generation for various resource-constrained microcontrollers. However, the analytics and ML part had to run in the cloud or any node in the distributed system that was not resource-constrained. Table \ref{tab:related-work} shows a comparison of the proposed approach with the related work in the literature.

\begin{table}[htbp]
	\caption{Relating the proposed approach to the previous work}
	\begin{center}
		\begin{tabular}{|p{1.75cm}|p{1.25cm}|p{1cm}|p{1.75cm}|p{1cm}|}
			\hline	
			\textbf{Approach} & \textbf{Basis} & \textbf{ML-enabled} & \textbf{Resource-constrained edge} & \textbf{TinyML} \\
			\hline
			ThingML \cite{ThingML} / HEADS \cite{HEADS} & EMF \cite{emf} \& Xtext \cite{xtext} & \cellcolor{red!25}- & \checkmark & \cellcolor{red!25}- \\
			\hline
		    $\mu$-Kevoree \cite{Fouquet+2012} & KMF \cite{kmf} & \cellcolor{red!25}- & \checkmark & \cellcolor{red!25}- \\
			\hline
			GreyCat \cite{GreyCat} & $\mu$-Kevoree \cite{Fouquet+2012} & \checkmark & \cellcolor{red!25}- & \cellcolor{red!25}- \\
			\hline
			ML-Quadrat \cite{ML-Quadrat} & ThingML \cite{ThingML} & \checkmark & \checkmark & \cellcolor{red!25}- \\
			\hline
			The proposed approach & ML-Quadrat \cite{ML-Quadrat} & \checkmark & \checkmark & \checkmark \\
			\hline
		\end{tabular}
		\label{tab:related-work}
	\end{center}
\end{table}

Finally, the proprietary software tooling provided by MathWorks\textsuperscript{\textregistered} Inc., which comprises MATLAB\textsuperscript{\textregistered}, Simulink\textsuperscript{\textregistered}, ThingSpeak\textsuperscript{\texttrademark} and other MATLAB\textsuperscript{\textregistered} add-ons, is widely used in the industry. However, the technology stack is based on the MATLAB\textsuperscript{\textregistered} ecosystem with full code generation in several programming languages, such as C/C++, HDL, .NET and Java, and with integrated APIs for relational and non-relational databases, as well as several communication protocols, e.g., REST, MQTT and OPC Unified Architecture (OPC UA) for both offline data (i.e., batch processing) and online data (i.e., stream processing). Note that the proprietary solution is expensive, thus a potential barrier for open innovations. In contrast, the solutions listed in Table \ref{tab:related-work} are open source with permissive licenses that enable their cost-effective deployment and possible future extensions. For instance, $\mu$-Kevoree \cite{Fouquet+2012} was the basis for GreyCat \cite{GreyCat}, ThingML \cite{ThingML} and HEADS \cite{HEADS} were the basis of ML-Quadrat \cite{ML-Quadrat}, and the present work builds on top of ML-Quadrat \cite{ML-Quadrat}. It is clear from the table that none of the prior work has addressed the TinyML platforms in its target IoT platforms for code generation.

\section{Proposed Approach}\label{proposed-approach}
\begin{figure}[htbp!]
	\centerline{\includegraphics[width=0.4\textwidth]{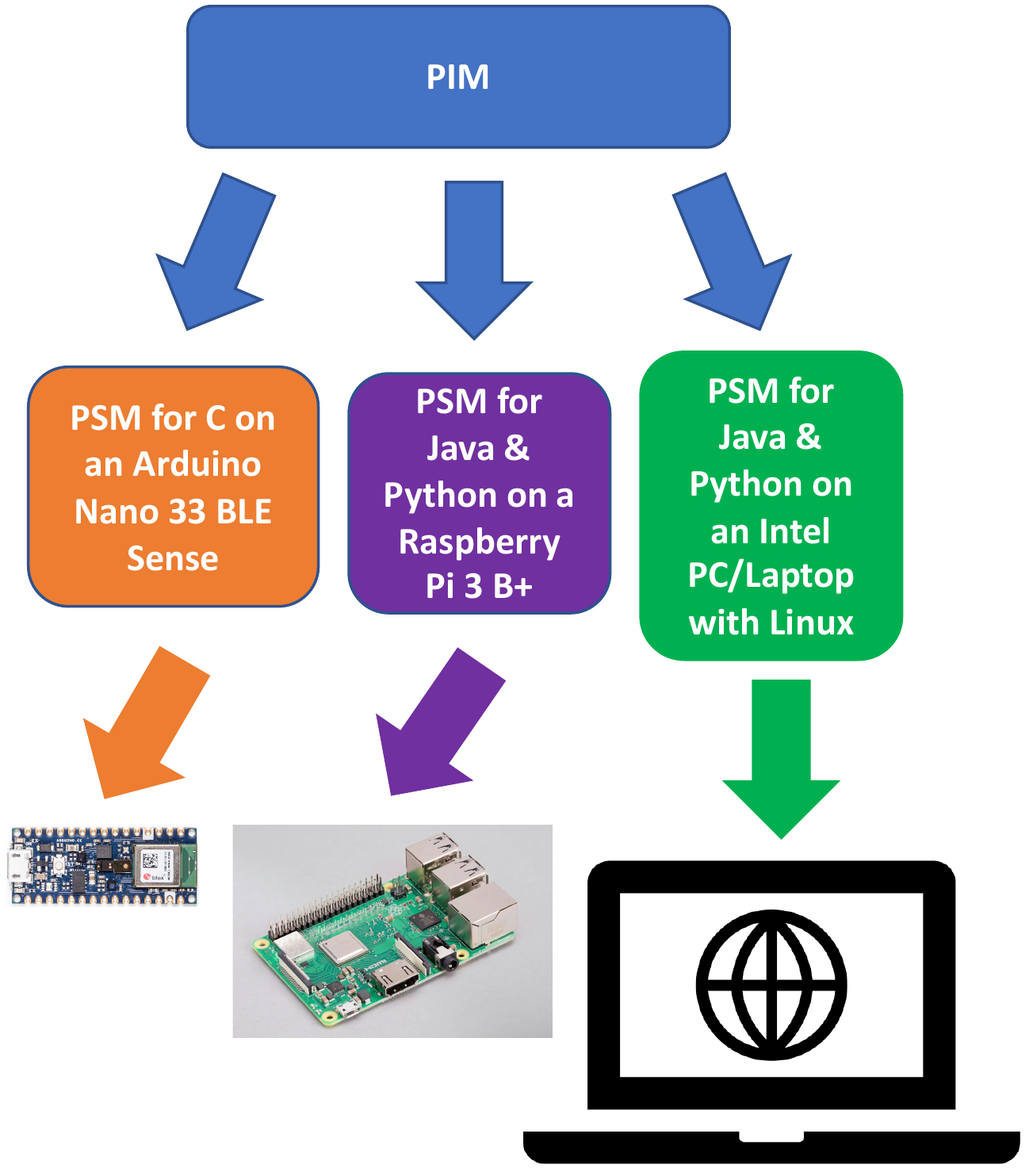}}
	\caption{From PIM to PSMs and full code generation. The images of Arduino and Raspberry Pi are from \cite{ArduinoNano33BleSense} and \cite{RaspberryPi3B+}, respectively.}
	\label{fig:pim-psm}
\end{figure}

We propose a novel approach to software and AI engineering for smart IoT services that enable their data analytics and ML parts to run in part or completely on the IoT edge devices that might be highly constrained in terms of their power and computational resources. The proposed approach is based on the DSM methodology \cite{KellyTolvanen2008} of the MDSE paradigm for software development. In particular, it builds on the prior work ML-Quadrat \cite{ML-Quadrat, Moin+2022-SoSyM}, thus integrates model-driven software and AI (specifically ML) engineering. As shown in Figure \ref{fig:pim-psm}, we generate code for heterogeneous IoT platforms out of the Platform-Specific Models (PSMs). However, we also offer generic models that are similar to the PIM viewpoint models in MDA \cite{OMGMDA2014, OMGCephas2006}. A PIM here can be considered as the \textit{common denominator} of the PSMs for a particular IoT service that must run on several heterogeneous IoT platforms. PIMs are at a higher level of abstraction than PSMs and specify the business logic of the IoT services regardless of the platform-specific details. Hence, the practitioner may concentrate on the overall, platform-independent structural and behavioral design of the software system architecture without any concerns about the possible lack of knowledge and skills in the diverse hardware, software, network and AI technologies that are used in the heterogeneous IoT edge devices and in the cloud.

Formally, we define a platform-independent software model for a smart IoT service as follows (see Equation \ref{eq:pim}):

\begin{equation}
  \label{eq:pim}
  \text{PIM} = (\Psi, M\textsubscript{ML}, B)
\end{equation}

\begin{figure}[htbp!]
	\centerline{\includegraphics[width=0.5\textwidth]{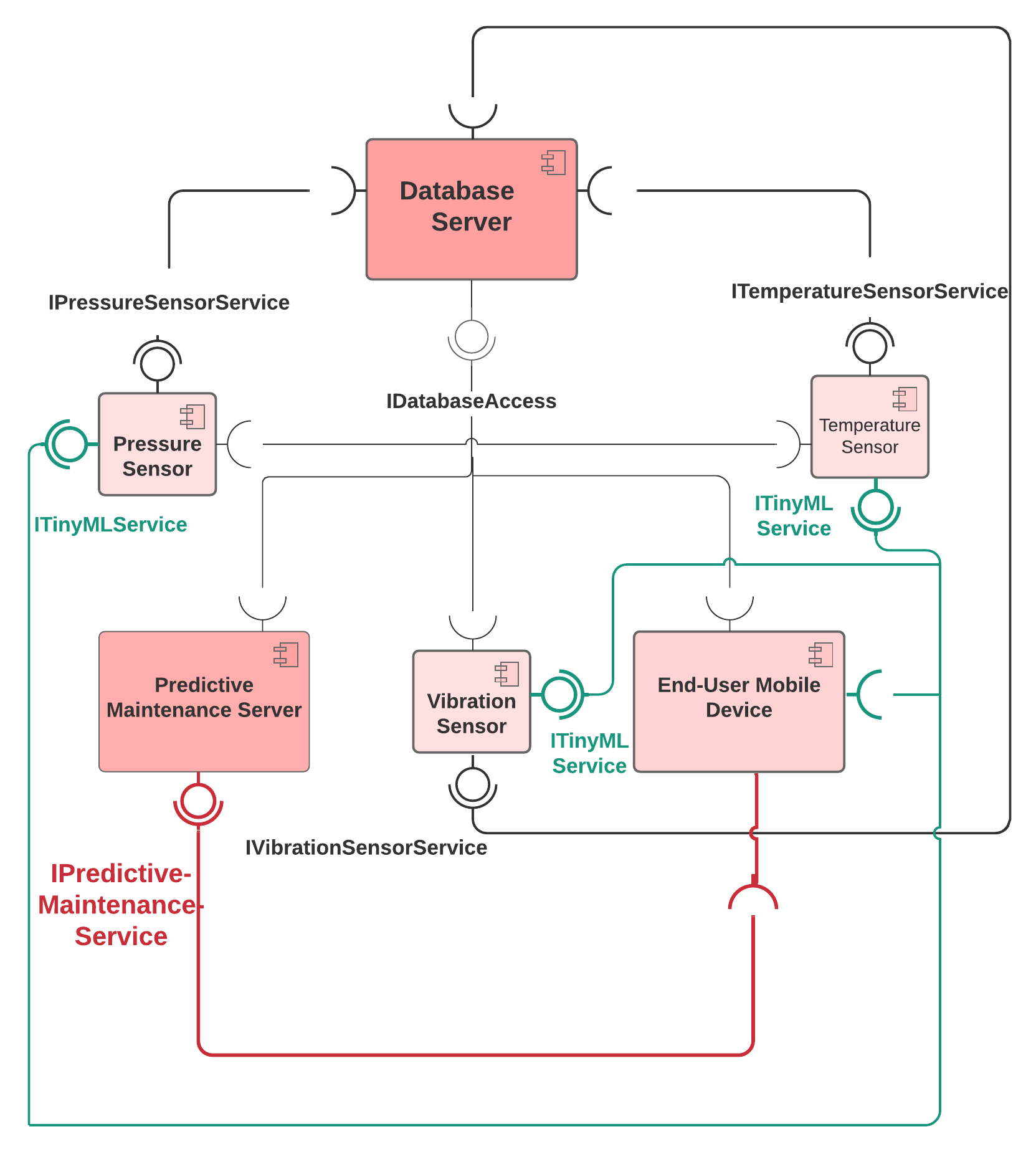}}
	\caption{The UML component diagram illustrating the structural architecture model of a sample IoT service for predictive maintenance.}
	\label{fig:component}
\end{figure}

\begin{figure}[htbp!]
	\centerline{\includegraphics[width=0.3\textwidth]{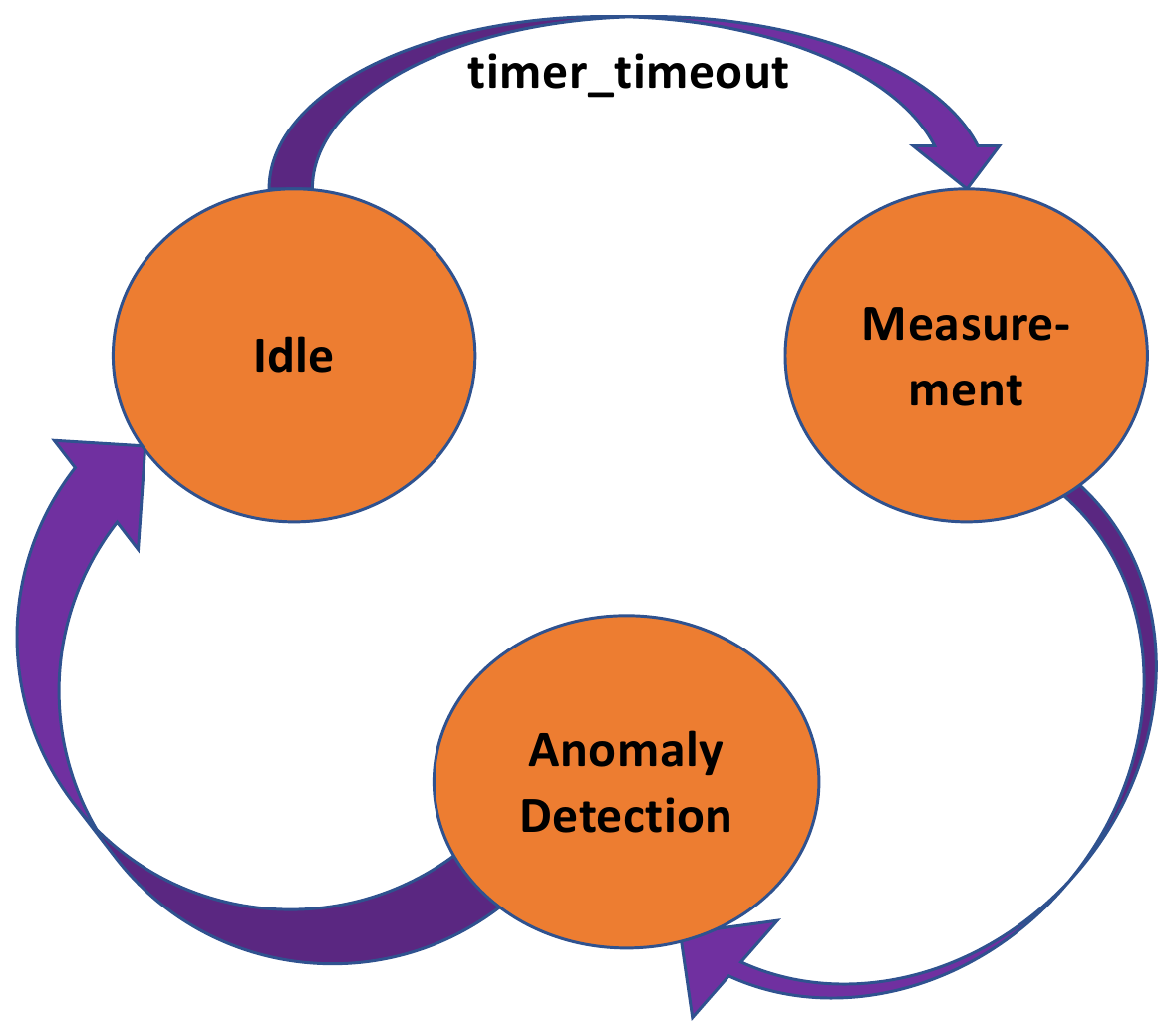}}
	\caption{The state machine diagram illustrating the behavioral architecture model of the sensors in the sample IoT service of Figure \ref{fig:component}.}
	\label{fig:statemachine}
\end{figure}

In Equation \ref{eq:pim}, $\Psi$ represents the structural elements of the software architecture model. Thus, it can be denoted, for example, by a component diagram. Figure \ref{fig:component} illustrates a sample UML component diagram for a predictive maintenance service. In addition, $M_{ML}$ is the ML model that brings AI to the IoT service. For instance, it can be an Artificial Neural Network (ANN), a Support Vector Machine (SVM) or a random forest ML model. Also, $B$ represents the behavioral elements of the software model\footnote{Note that $M_{ML}$ might affect $B$.}. Hence, it can be denoted, for example, by a state machine diagram. Figure \ref{fig:statemachine} demonstrates a sample state machine or statechart for the behavioral model of an IoT sensor that can also conduct anomaly detection via ML as its \textit{TinyML service}, in addition to its measurement service.

Furthermore, we define a platform-specific software model for a smart IoT service as in Equation \ref{eq:psm} below:

\begin{equation}
  \label{eq:psm}
  \text{PSM} = (\text{PIM}, A, C)
\end{equation}

\begin{figure}[htbp!]
	\centerline{\includegraphics[width=0.5\textwidth]{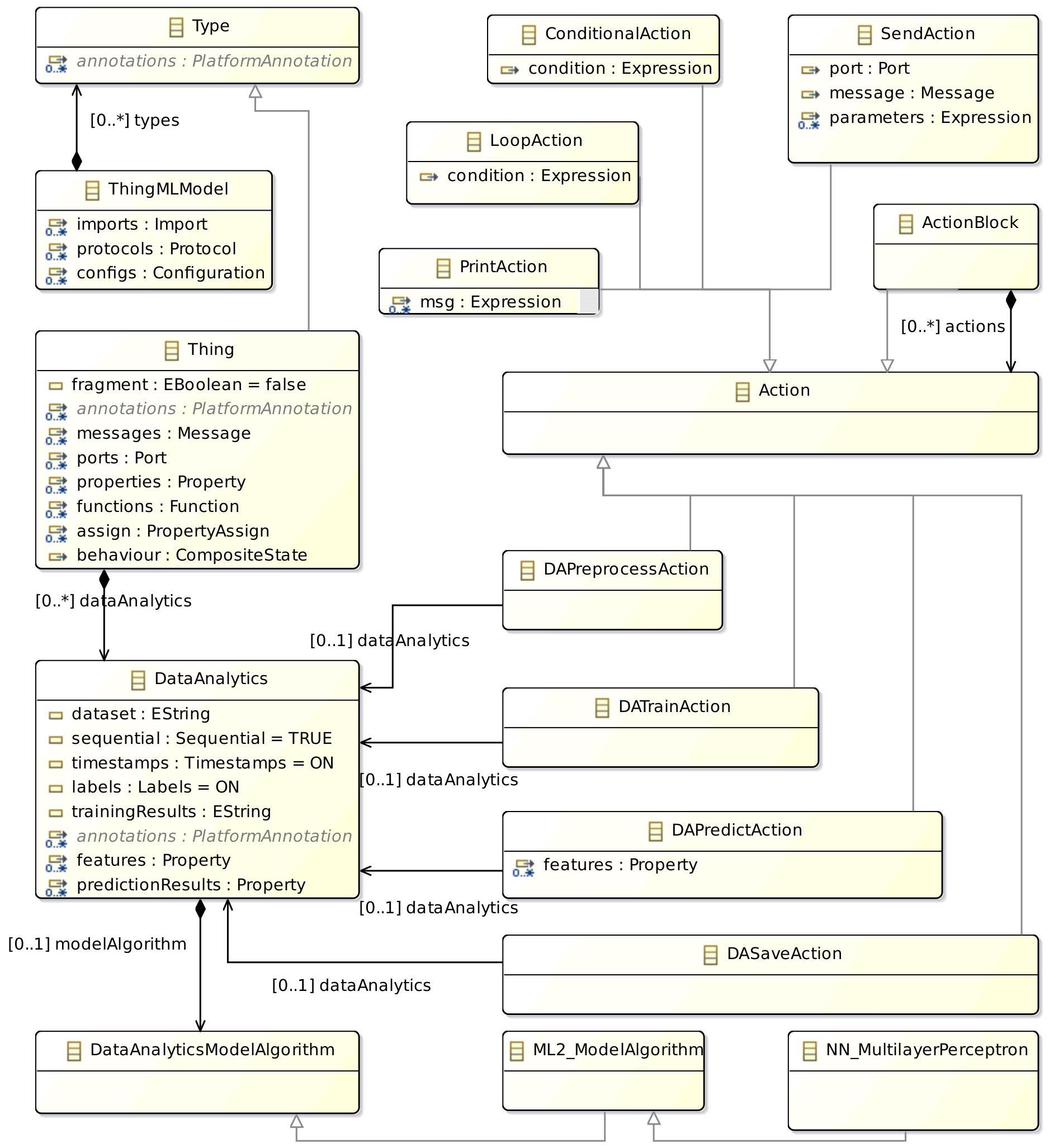}}
	\caption{Part of the meta-model that is adopted from ML-Quadrat \cite{Moin+2022-SoSyM}.}
	\label{fig:MM}
\end{figure}

Here, $A$ and $C$ stand for the platform-specific annotations and configurations, respectively. \textit{Annotations} can be attached to various elements of model instances to add platform-specific details. For instance, they may help model-to-code transformations (code generators) in mapping the data types to the right ones in the target platforms. This means, they can, for example, specify whether an Integer data type in the software model instance must be mapped to the \textit{int}, \textit{short} or \textit{long} type in the generated Java code. Furthermore, \textit{configurations} are required in order to make model instances ready for code generation out of them. Configurations must include object instances of the \textit{thing} classes and specify their interconnections for message-passing. In addition, they may include annotations that specify the target platform for code generation. In order to create a PSM out of a PIM, one must insert annotations and append configurations to the PIM for the respective target platform for code generation. We show this in Section \ref{validation}. Note that currently both PIM and PSM instances conform to the same meta-model that is adopted from ML-Quadrat \cite{ML-Quadrat, Moin+2022-SoSyM}. Figure \ref{fig:MM} illustrates part of this meta-model.

Moreover, we provide model-to-code transformations (i.e., code generators) that can produce the entire software solution for the desired smart IoT services (see the orange, purple and green arrows in Figure \ref{fig:pim-psm}). In the present work, we implement the code generator for the TinyML part that can generate the APIs of \textit{TensorFlow Lite} \cite{tensorflow-lite} and \textit{TensorFlow Lite for Microcontrollers} \cite{tensorflow-lite-mc}. For the validation case study in Section~\ref{validation}, we deploy the former on a \textit{Raspberry Pi 3 B+} \cite{RaspberryPi3B+} and the latter on an \textit{Arduino Nano 33 BLE Sense} microcontroller \cite{ArduinoNano33BleSense}. In addition, other code generators from prior work, e.g., ML-Quadrat \cite{ML-Quadrat, Moin+2022-SoSyM} can be used to generate code for a Linux PC/laptop with an Intel x86 CPU. In this case, the Keras \cite{Chollet+2015} API for TensorFlow \cite{Abadi+2015} will be generated and used for the ML part.

The proposed approach enables deploying pre-trained ML models on various distributed edge devices without sharing data. This means, ML models, such as ANNs can be created and trained with existing data on a server, e.g., on-premises or in the cloud. Then, depending on the size of the trained ML models, the requirements and the capabilities of the IoT devices, some of the ML models may be deployed on TinyML platforms, and some other ML models may be deployed on other edge devices, such as smartphones, smart home appliances and gateways, or in the cloud. One key benefit of bringing the data analytics and ML models to the edge of the network is the ability to respect the possible privacy and security concerns or regulations. Instead of asking the users to upload their new data to the cloud and making predictions there, we let the users keep their new data on their side and enable predictions there.

In the context of the predictive maintenance use case that is illustrated in Figure \ref{fig:component} and Figure \ref{fig:statemachine}, one might think of $m+1$ ML models to be deployed on $m$ sensors and on one server or controller. The latter will be more capable in terms of the computational power, thus can run a more advanced ML model and take more data into account, whereas the former will be restricted ML models for carrying out simpler tasks on a local level. In other words, each sensor might, for example, process the data that come from its own measurements and possibly its neighbors in the sensor network, whereas the server or controller node that must be more capable conducts a more extensive condition-based monitoring of the entire system. This means, the predictive maintenance operations can be executed on the local, i.e., sensor and global levels. These are shown via the green and the red lines in Figure \ref{fig:component}, respectively. This \textit{federated} design\footnote{Here, federated should not be confused with the notion of \textit{federated learning} \cite{Li+2021} in which an ML model is built collaboratively in a distributed system without sharing data among the participants.} is expected to increase the availability of the services, e.g., condition-based monitoring, and contribute to the fault tolerance and the overall resilience of the system since multiple nodes will perform the ML task independently and with different qualities according to their resources. If a network or power outage occurs in one part of the system, other nodes can still deliver some level of service. 

In addition, the network throughput might be reduced by letting individual sensors perform some basic analytics tasks locally, thus reducing the frequency and the amount of the data that needs to be sent to the database and/or the server/controller. Further, in certain applications, the TinyML operations on the resource-constrained nodes might suffice, thus resulting in a much lower level of energy consumption for the data analytics and ML tasks. This should lead to environmental care and sustainability in the long term. Finally, by deploying ML models on the TinyML devices with ultra-low power consumption, AI/ML can become more affordable, and can also be brought to the situations where Internet connectivity is not possible and/or to the extreme conditions where sensors that will be mounted somewhere, e.g., under the ground, in the oceans or on very large structures must run on their limited battery powers for a relatively long time and are not physically accessible in a cost-effective manner after their initial installation.

\section{Implementation \& Validation}\label{validation}
We implement the proposed approach by extending the ML-Quadrat \cite{ML-Quadrat} project. First, we adopt the Xtext-based meta-model (grammar) of ML-Quadrat \cite{ML-Quadrat}. Second, to support the new target platforms, namely Raspberry Pi 3 B+ \cite{RaspberryPi3B+} and Arduino Nano 33 BLE Sense \cite{ArduinoNano33BleSense}, we introduce new annotation types for \textit{configurations} that enable practitioners to choose the said platforms as target platforms for code generation. We extend the model-to-code transformations, i.e., the code generators of ML-Quadrat \cite{ML-Quadrat} to support full code generation for the said platforms. To this aim, we deploy the APIs of the TensorFlow Lite \cite{tensorflow-lite} and TensorFlow Lite for microcontrollers \cite{tensorflow-lite-mc} libraries. In the former case, the generated code is in Python, whereas in the latter case it is C code for Arduino. The code generation process also includes the conversion of ML models to the proper formats that are acceptable by the mentioned libraries. In the latter case, namely the microcontroller, this format is a hexadecimal dump of a C array that is stored in a C source file. Further, the code generators themselves are implemented in Java.

In the following, we validate the proposed approach through a case study. There exists a hydraulic test rig that is deployed in Saarbr{\"u}cken, Germany \cite{Helwig+2015}. A test rig or test station is used to test and assess the capability and performance of components for industrial use \cite{hydrotechnik}. The hydraulic test rig is equipped with multiple sensors and the sensor data, as well as the data about the working conditions and status of the system is provided by the ZeMA gGmbH research center for Mechatronics and automation technologies as open data \cite{Kaggle-Hydraulic}. We use the data from the following 3 sensors in order to predict any possible internal leakage of the main pump: (i) The vibration sensor (VS1) of the main pump. Its readings are recorded in the mm/s unit and at the frequency of 1 Hz (i.e., once a second). (ii) The Electrical Power Sensor (EPS1) of the main pump. Its readings are recorded in Watts and at the frequency of 100 Hz. (iii) The System Efficiency (SE) factor that is not a real (i.e., physical) sensor, but a virtual sensor. Its values are determined by combining different directly measured values \cite{Helwig+2015}. Moreover, it is a percentage and has the frequency or sampling rate of 1 Hz. Since the hydraulic test rig repeats periodic constant load cycles of 60 seconds, we require the sensor data for one cycle in order to predict whether the main pump is prone to any internal leakage or not. We let an Artificial Neural Network (ANN) model perform this prediction. The ML features that are used to train this model are the above-mentioned sensor values, namely VS1, EPS1 and SE. As the sampling rates or frequencies are not identical, one could, for example, down-sample EPS1 that has a frequency of 100 Hz to 1 Hz. However, for the current use case, we keep it as it is. Therefore, there exist 60 features for VS1, 6,000 features for EPS1 and another 60 features for SE per system cycle. We assign one Boolean/Binary class label to each cycle that is either True (i.e., leakage positive) or False (i.e., leakage negative). The dataset contains 2,205 data instances, i.e., system cycles. Out of these 2,205 instances, 1,221 instances/cycles (55\%) correspond to no leakage and the rest corresponds to leakage.

Since we are dealing with time series data in which the order of the data instances matters, we avoid shuffling the data samples/instances. We separate the dataset into two parts. We dedicate 80\% of the available data to the training and validation dataset and the rest to the test dataset for the evaluation. The latter must remain unseen by the ML model to make a fair evaluation possible. The choice of 80\% vs. 20\% is a common practice in ML and also aligns with the Pareto principle that is widely used in science and engineering.

\begin{figure}[htbp!]
	\centerline{\includegraphics[width=0.5\textwidth]{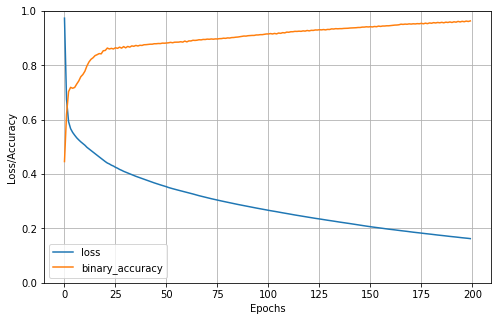}}
	\caption{The values of the loss function and the binary accuracy during the training of the ML model.}
	\label{fig:training-loss}
\end{figure}

We standardize the numeric training data using Z-Scores and train an Artificial Neural Network (ANN) model using the data. It transpires that a \textit{Multi-Layer Perceptron (MLP)} with three layers (input, hidden and output) accomplishes the prediction task with high accuracy, precision and recall. The hidden layer is a \textit{Dense} layer with 32 units in the first experiment and 8 units in the second experiment below, as well as the \textit{Relu} activation function. Further, the output layer has 2 units and the \textit{Sigmoid} activation function. Moreover, we use the \textit{Adam} optimizer, the \textit{Binary Crossentropy} loss function, a learning-rate of \textit{1e-5}, a batch size of 100, 200 training epochs, as well as early-stopping with a patience level of 3. Figure \ref{fig:training-loss} depicts the changes of the loss function and the binary accuracy during the training of the ML model.

We could deploy more complex and advanced ANN architectures, e.g., Recurrent Neural Networks (RNN), such as Long-Short Term Memory (LSTM) layers. However, since the mentioned architecture already performs well and we prefer a more compact ML model, we leave it like this. According to the experimental results on the test dataset that are illustrated in Table \ref{tab:exp-results}, the accuracy, precision and recall were $97\%$, $97\%$ and $97\%$, respectively, for the first experiment (i.e., with 32 units in the hidden layer of the above-mentioned ML model), and $80\%$, $86\%$ and $80\%$, respectively, for the second experiment (i.e., with 8 units in the hidden layer of the above-mentioned ML model) on an Intel x86 platform with the Linux Operating System (OS) and Python code that deploys the TensorFlow \cite{Abadi+2015} library. This Linux server has 45GB of main memory (RAM) and 10 Intel Xeon 2.3 GHz Processors. The respective rows in Table \ref{tab:exp-results} are colored in gray. As shown, the more compact ML model (namely the latter experiment) performs faster. Thus, it requires only $86$ milliseconds for the entire test dataset instead of $119$ milliseconds in the first experiment (i.e., $38\%$ time reduction). However, the accuracy and recall have been reduced by $17.5\%$ each, and the precision has fallen by $11.3\%$ in the second experiment with the more compact ML model that is $74.5\%$ smaller in size.

Further, the second and the sixth rows in Table \ref{tab:exp-results} demonstrate the experimental results for the first and the second experiment on a Raspberry Pi (RPI) platform, respectively. As mentioned, this is a Raspberry Pi 3 B+ \cite{RaspberryPi3B+} board with the Raspberry Pi OS (formerly known as Raspbian) and Python code that deploys the TensorFlow Lite \cite{tensorflow-lite} library. This library provides an API for an ML model converter that enables generating an optimized ML model in the FlatBuffers \cite{flatbuffers} serialization format and the \textit{.tflite} file extension \cite{tensorflow-lite}. As we can see in the table, this conversion results in $67\%$ and $68\%$ ML model size reduction in the first and the second experiments, respectively, without compromising the ML model performance in terms of its accuracy, precision and recall. Nevertheless, it is clear that predictions on the RPI platform need more time, due to the limitations of the computational resources, compared to the said Linux server. The increases in the prediction time for the entire test dataset are $1,100\%$ and $694\%$ for the first and the second experiments, respectively.

In addition, we apply a technique, called \textit{post-training quantization} \cite{tensorflow-lite} in both experiments and illustrate the results in the third and the seventh rows in Table \ref{tab:exp-results}. Hence, with a negligible compromise in the ML model performance in terms of accuracy, precision and recall, we reduce the ML model size considerably and speed up its predictions too. For instance, in the case of our first experiment on the RPI platform, we do not face any reduction in the accuracy, precision or recall. Also, in the second experiment, the accuracy and recall remain the same, while the precision is reduced by only $1\%$. However, the quantization technique results in an ML model size reduction of $75\%$ and $74\%$ in the first and the second experiments, respectively. Note that this quantized ML model makes predictions $60\%$ and $29\%$ faster in the first and the second experiments, respectively. In this case, quantization leads to converting all of the \textit{float32} weights of the ML model to \textit{int8} values.

While both the non-quantized and the quantized variants of the above-mentioned ML model fit into the main memory of the RPI board, for the TinyML platform, namely the Arduino Nano 33 BLE Sense microcontroller \cite{ArduinoNano33BleSense}, the situation is different. To deploy the ML model on this platform, we use the \textit{xxd} Unix/Linux command to generate a hexadecimal dump of the mentioned FlatBuffers model as a C Byte Array. We store the resulting ML model in a C++ source file with the \textit{.cc} extension. This can be used via the TensorFlow Lite for microcontrollers \cite{tensorflow-lite-mc} library on the Arduino microcontroller. However, the main issue is that the hexadecimal dump requires more space on the disk than the efficient FlatBuffers serialization. Note that in the case of the first experiment, the C Byte Array has a size of 198 KB (i.e., with 198 thousands elements). However, its hexadecimal dump requires 1.2 MB disk space. This is $506\%$ larger. Since the main memory of the microcontroller has only 1 MB, which is even large compared to many TinyML platforms, we cannot deploy this ML model on the Arduino platform. In fact, this is the reason that we conduct the second experiment with a more compact ML model. Here, we have a C Byte Array of 51 KB (i.e., with 51 thousands elements). Again, the hexadecimal dump requires more space. In this case, it takes 316 KB on the disk. Therefore, it can fit into the main memory of the TinyML platform. The results of both experiments on Arduino are shown on the fourth and the eights rows of Table \ref{tab:exp-results}.

Recall from Section \ref{introduction} that we have 2 Research Questions (RQ): \textbf{RQ1.} Can we enable automated full code generation out of the software models of smart IoT services that will deploy trained ML models on highly resource-constrained IoT edge platforms (i.e., realizing TinyML)? \textbf{RQ2.} Can we have a higher level of abstraction for the Platform-Independent Models (PIM) that will abstract from the details and constraints of the underlying IoT platforms, and simultaneously a lower level of abstraction for the Platform-Specific Models (PSM) out of which the full implementation must be generated?

\begin{figure}[htbp!]
	\centerline{\includegraphics[width=0.5\textwidth]{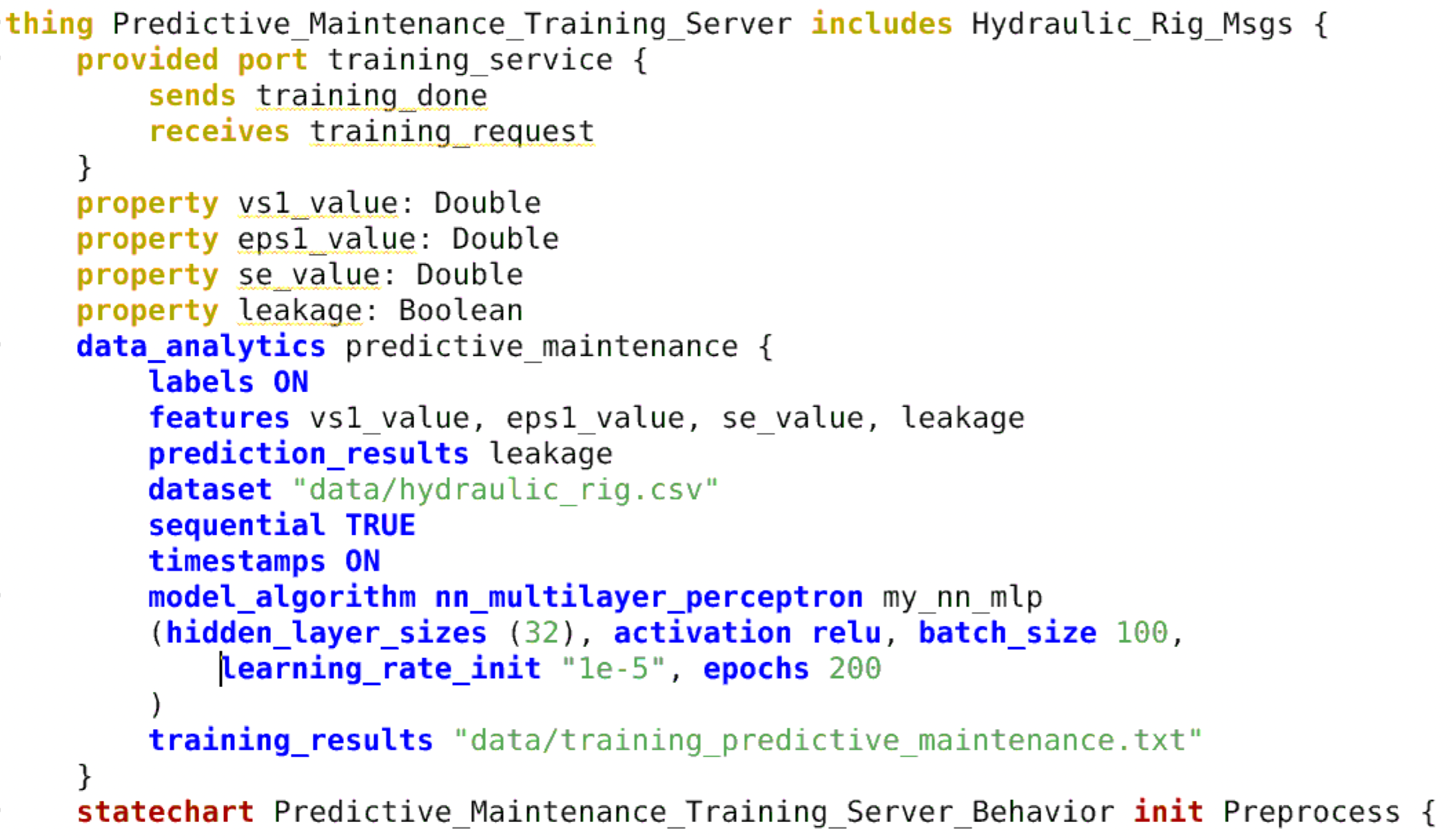}}
	\caption{Part of the PIM for training the ML model.}
	\label{fig:training-pim}
\end{figure}

\begin{figure}[htbp!]
	\centerline{\includegraphics[width=0.5\textwidth]{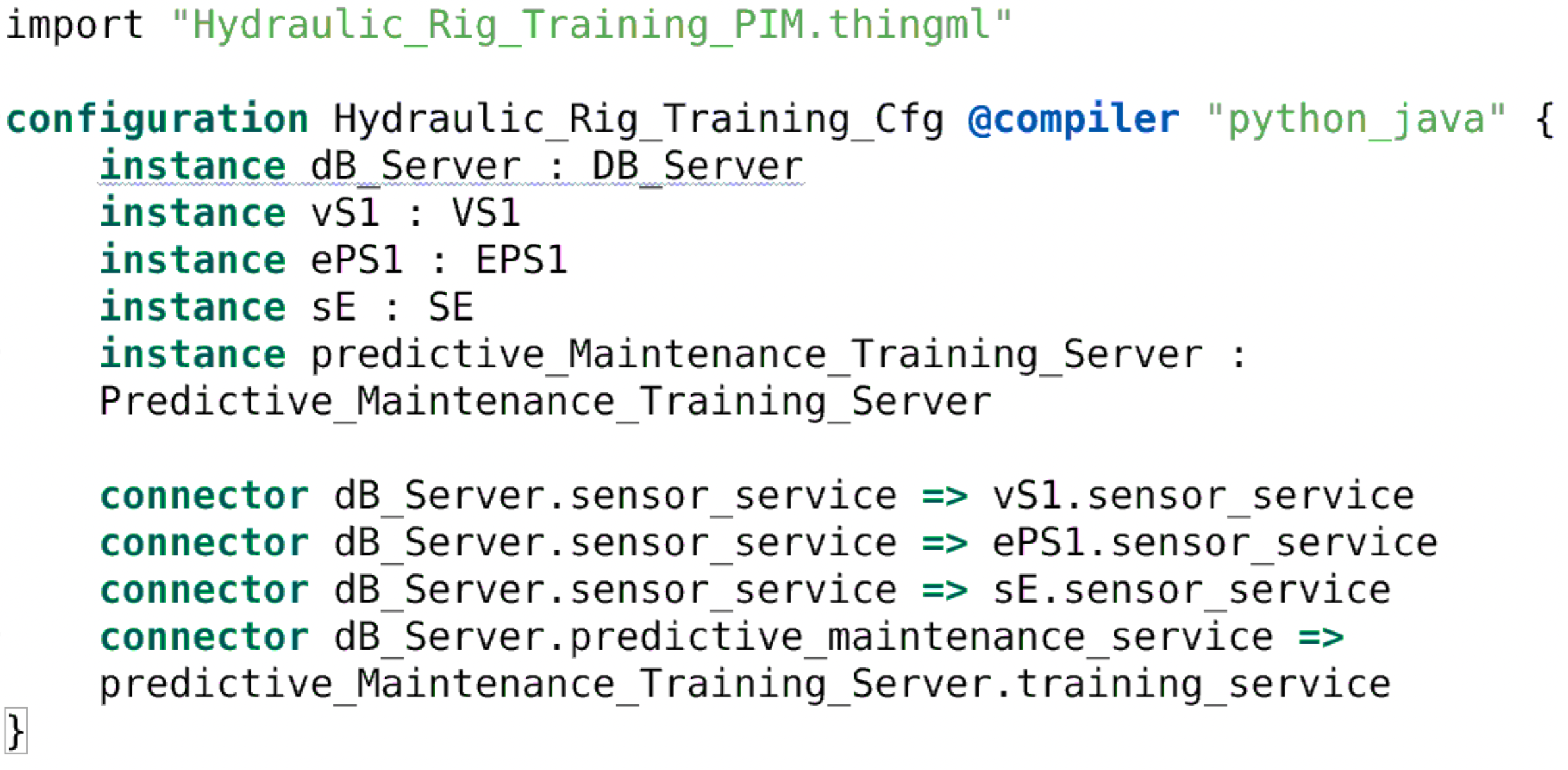}}
	\caption{Part of the PSM for training the ML model on an x86 Linux platform. The PIM of Figure \ref{fig:training-pim} is imported here.}
	\label{fig:training-psm}
\end{figure}

To assess and validate the research questions, we realize the above-mentioned use case with our textual Domain-Specific Modeling Language (DSML) that is based on ML-Quadrat \cite{ML-Quadrat} and let the full source code become generated automatically out of the software model instances. The generated code can create, train and deploy the said ML models. To this aim, we implement both PIMs and PSMs. In fact, we need two PIMs and two PSMs for training the ML model on the x86 Linux machine. The two PIMs are responsible for creating and training the ML models of the two experiments, respectively (namely, rows 1-4 and rows 5-8 in Table \ref{tab:exp-results}). Here, we illustrate the PIM for the first experiment. Figure \ref{fig:training-pim} and Figure \ref{fig:training-psm} show part of the platform-independent and platform-specific software model instances for the training on the x86 Linux server. If we wanted to train the ML model on another platform, such as a Raspberry Pi, we would take the same PIM and import it in another PSM that would have been specific to that platform. The choice of the target platform for code generation is specified through the \textit{@compiler} annotation of the \textit{configurations} (see Figure \ref{fig:training-psm}). For instance, \textit{@compiler python\_java} generates Python and Java code for the default platform, namely an X86 Linux machine. The other PIM is also similar. However, the value of the parameter \textit{hidden\_layer\_sizes} is $8$ rather than $32$. 

In addition, we require two PIMs for making predictions using the ML models. Again, the two PIMs correspond to the two experiments (namely, rows 1-4 and rows 5-8 in Table \ref{tab:exp-results}). These two PIMs for prediction are imported in the PSMs for prediction. In principle, we need one PSM for each of the target platforms. Since we have three target platforms (x86, RPI and Arduino) and two experiments, we have in total $6$ PSMs for prediction. We depict part of one of the PSMs for prediction on the x86 platform in Figure \ref{fig:prediction-psm}. The other ones are also similar to this one. However, their \textit{@compiler} annotations in their \textit{configurations} have the values \textit{rpi\_3b+\_python}, \textit{rpi\_3b+\_python\_quantized} and \textit{arduino\_nano\_33\_ble\_sense\_cpp} for each of the respective target platforms. The code generators not only produce the source code that has the APIs of these platforms, but also automatically converts the ML model to the right format for each of the specific platforms.

\begin{figure}[htbp!]
	\centerline{\includegraphics[width=0.25\textwidth]{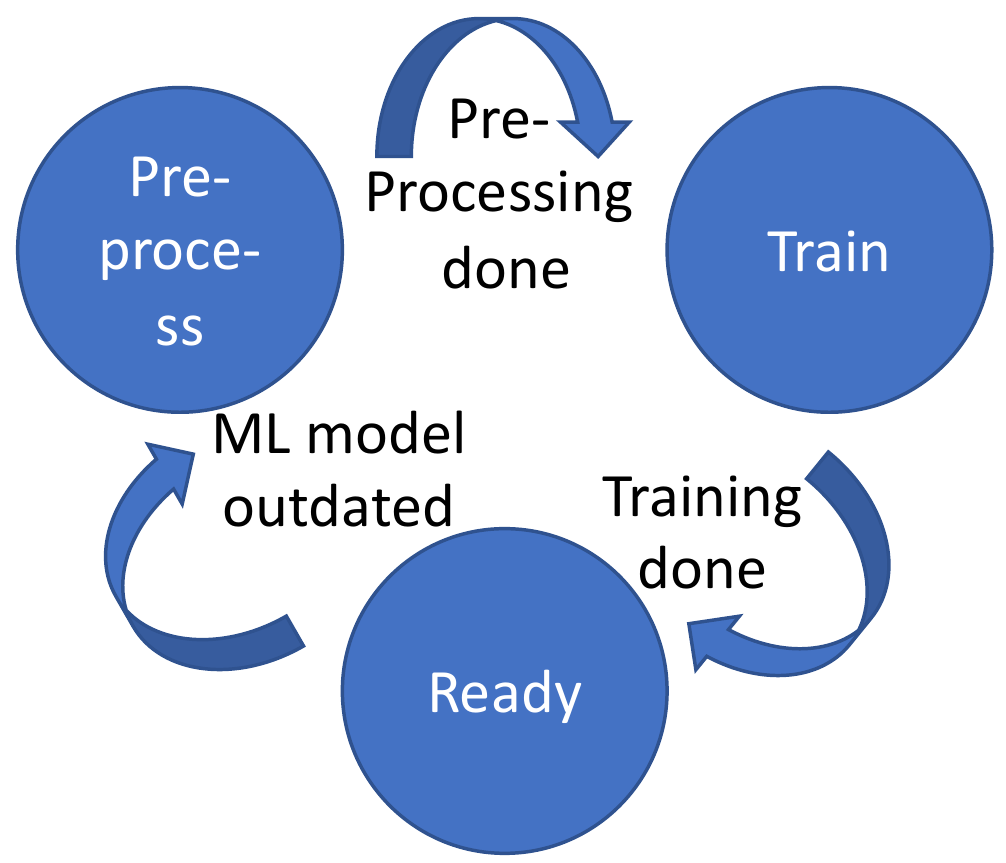}}
	\caption{The behavioral model of the PIM for training the ML model.}
	\label{fig:FSM_Training}
\end{figure}

\begin{figure}[htbp!]
	\centerline{\includegraphics[width=0.5\textwidth]{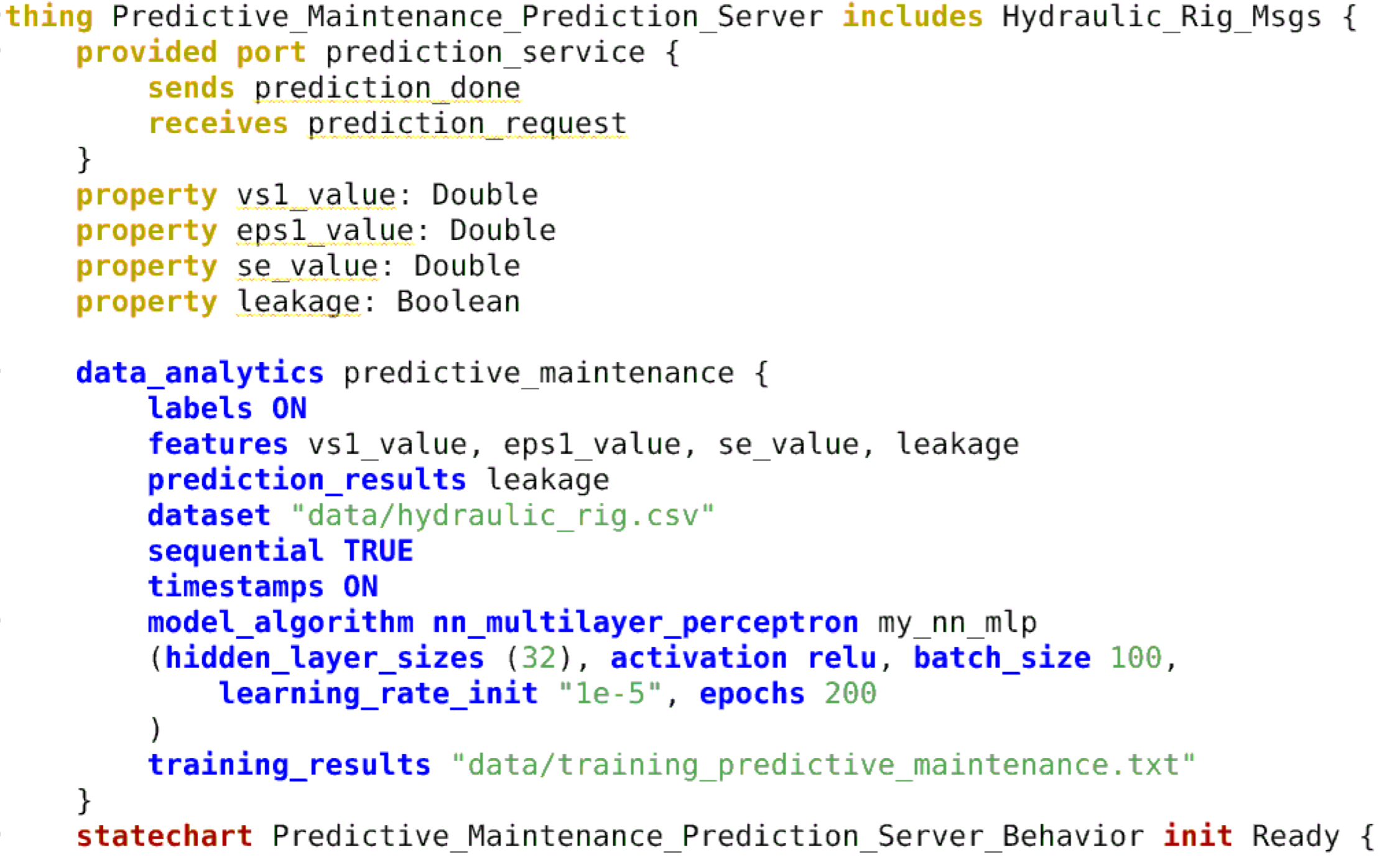}}
	\caption{Part of the PIM for predictions using the ML model.}
	\label{fig:prediction-pim}
\end{figure}

\begin{figure}[htbp!]
	\centerline{\includegraphics[width=0.25\textwidth]{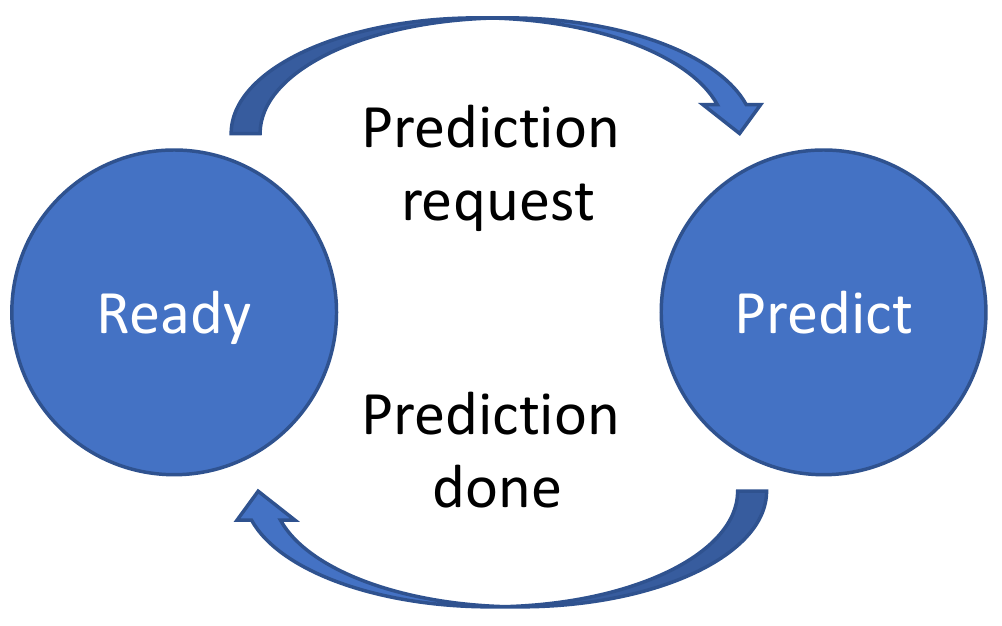}}
	\caption{The behavioral model of the PIM for predictions using the ML model.}
	\label{fig:FSM_Prediction}
\end{figure}

\begin{figure}[htbp!]
	\centerline{\includegraphics[width=0.5\textwidth]{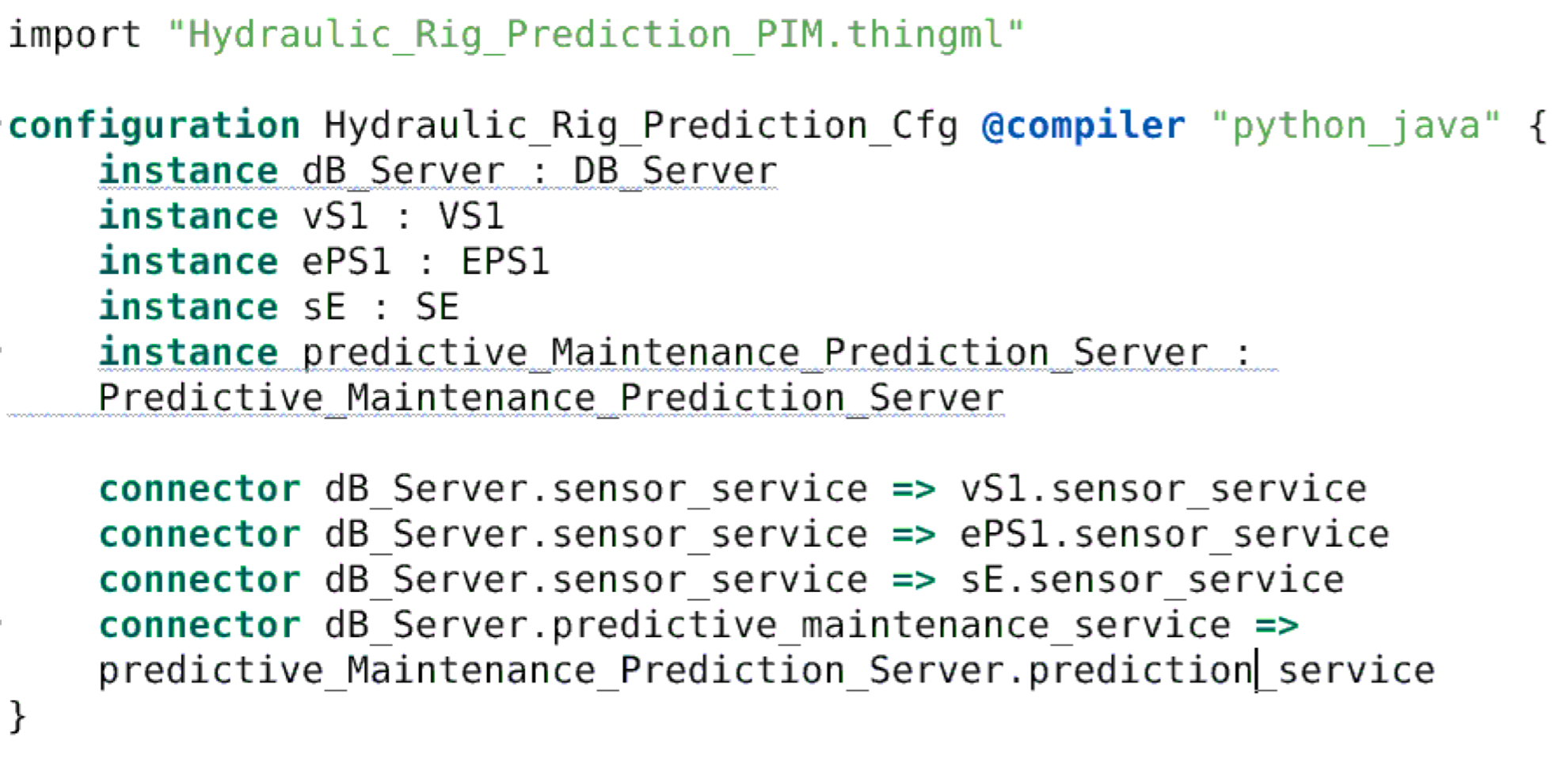}}
	\caption{Part of the PSM for predictions using the ML model on an x86 Linux platform. The PIM of Figure \ref{fig:prediction-pim} is imported here.}
	\label{fig:prediction-psm}
\end{figure}

To make it more comprehensible, we show the behavioral part of the model instances of Figure \ref{fig:training-pim} and Figure \ref{fig:prediction-pim} in the graphical form in Figure \ref{fig:FSM_Training} and \ref{fig:FSM_Prediction}, respectively. The implementation of the state machines is carried out through the \textit{statechart} section of the textual model instances (see the last lines in Figure \ref{fig:training-pim} and Figure \ref{fig:prediction-pim}).

For more information about the syntax of ML-Quadrat, please refer to its documentation \cite{ML-Quadrat} and prior work \cite{Moin+2022-SoSyM}.

\begin{table}[htbp]
	\caption{Experimental results}
	\begin{center}
		\begin{tabular}{|p{0.1cm}|p{1.5cm}|p{0.7cm}|p{0.7cm}|p{0.7cm}|p{0.7cm}|p{1.25cm}|}
			\hline	
			& \textbf{Experi- ment \& Platform} & \textbf{Predic- tion time (s)} & \textbf{Accu- racy} & \textbf{Prec- ision} & \textbf{Rec- all} & \textbf{ML Model size} \\
			\hline
			\cellcolor{gray!25}1 & \cellcolor{gray!25}1, x86 Linux & \cellcolor{gray!25}$0.119$ & \cellcolor{gray!25}$97\%$ & \cellcolor{gray!25}$97\%$ & \cellcolor{gray!25}$97\%$ & \cellcolor{gray!25}$2.4$ MB \\
			\hline
		    2 & 1, RPI & $1.43$ & $97\%$ & $97\%$ & $97\%$ & $785$ KB \\
			\hline
			3 & 1, RPI, Q. & $0.572$ & $97\%$ & $97\%$ & $97\%$ & $198$ KB \\
			\hline
			4 & 1, Ard., Q. & n/a & n/a & n/a & n/a & \cellcolor{red!75}$1.2$ MB \\
			\hline
			\cellcolor{gray!25}5 & \cellcolor{gray!25}2, x86 Linux & \cellcolor{gray!25}$0.086$ & \cellcolor{gray!25}$80\%$ & \cellcolor{gray!25}$86\%$ & \cellcolor{gray!25}$80\%$ & \cellcolor{gray!25}$613$ KB \\
			\hline
		    6 & 2, RPI & $0.683$ & $80\%$ & $86\%$ & $80\%$ & $197$ KB \\
			\hline
			7 & 2, RPI, Q. & $0.482$ & $80\%$ & $85\%$ & $80\%$ & $51$ KB \\
			\hline
			8 & 2, Ard., Q. & $218$ & $80\%$ & $85\%$ & $80\%$ & \cellcolor{green!75}$316$ KB \\
			\hline
		\end{tabular}
		\label{tab:exp-results}
	\end{center}
\end{table}

\section{Conclusion \& Future Work}\label{conclusion-future-work}
In this paper, we proposed a novel approach to model-driven development of smart IoT services that can be deployed on a wide variety of distributed platforms, including the highly resource-constrained, ultra low-power microcontrollers. We enabled TinyML, thus supported the deployment of compact ML models on the said microcontrollers. To this aim, we used the APIs of the TensorFlow Lite \cite{tensorflow-lite} and the TensorFlow Lite for microcontrollers \cite{tensorflow-lite-mc} libraries.

First, we validated RQ1 by showing the feasibility of full source code generation in an automated manner. In addition to generating the source code for the different target platforms, we also automatically converted the ML models to the right formats for each of them. Second, we validated RQ2 concerning the different levels of abstraction on the modeling layer: PIMs vs. PSMs. We support importing a PIM that abstrcats from the underlying platform-specific details in multiple PSMs, such that the PSMs can add the APIs of the target platforms and enable full code generation out of them. 

The validation was performed through a case study. While this is a common empirical research method, we acknowledge that this single case study and use case scenario might not be representative enough for the entire IoT domain. Therefore, future research work is required to assess and validate the proposed approach for multiple other scenarios and cases. In addition, we used the open data of a hydraulic test rig for the validation. Concerning the reported evaluation results, we must note that any dataset typically contains a certain degree of noise and often has multiple missing values. In this work, we did not handle any imputation of missing values since the provided dataset did not contain any. We assume that the data have already been cleaned before being released publicly.

Furthermore, future work can extend the proposed approach to enable federated learning by embedded platforms and TinyML devices such that an ML model can be built collaboratively without sharing raw data between the devices. In contrast, in the present work, we trained the ML model on one platform (x86 Linux) and deployed it on three different platforms, including a TinyML platform.

%\section*{Data Availability}
%The authors are committed to the open science initiative. Therefore, the entire research data are available as open data under the terms of the Creative Commons Attribution 4.0 International license at \url{https://doi.org/10.5281/zenodo.5426130}.

%\begin{comment}
\appendix
\section{Appendix}\label{appendix}
In the following, we briefly explain some of the keywords of the textual concrete syntax of the DSML of ML-Quadrat \cite{ML-Quadrat} that is adopted in this work:

\textbf{Labels:} This is a binary parameter that can have the values ON, OFF or SEMI for supervised, unsupervised and semi-supervised ML, respectively.

\textbf{Features:} This is a list of the properties (i.e., local variables) of the \textit{thing} that must be considered as ML features (attributes). The local variables might include the messages or parameters of the messages.

\textbf{Prediction\_results:} This parameter determines the property of the \textit{thing} in which the prediction result of the ML model must be stored.

\textbf{Sequential:} A Boolean parameter that indicates whether the input data are sequential, e.g., time series, where the order of data instances matters. In this case, shuffling and cross-validation should be avoided.
 
\textbf{Timestamps:} A binary parameter that states if the data instances have timestamps.

\textbf{Model\_algorithm:} Here, one can specify the particular ML model architecture that must be deployed, e.g., the Multi-Layer Perceptron (MLP) Neural Networks (NN). Additionally, the hyperparameters, e.g., the choice of the error/loss function, the learning/optimization algorithm, the learning rate, etc. might be given in the parenthesis. Each family of ML models may have a different set of possible hyperparameters.

\textbf{Training\_results:} This is the path of the text file in which the log of training must be stored.

%\end{comment}

\section*{Acknowledgments}
This work is partially funded by the German Federal Ministry for Education \& Research (BMBF) through the Software Campus initiative (project ML-Quadrat).

\section*{Notice}
This work has been submitted to the IEEE for possible publication. Copyright may be transferred without notice, after which this version may no longer be accessible.

\bibliographystyle{IEEEtran}
\bibliography{refs}

\end{document}